\documentstyle[12pt]{article}
\begin{document}
\pagestyle{empty}
\phantom{.}
\def\dpartial{\mbox{$\boldmath{\partial}$}}
\def\dgamma{\mbox{$\boldmath{\gamma}$}}
\def\dtau{\mbox{$\boldmath{\tau}$}}
\def\thesection{\Roman{section}}

{\hfill  \bf Preprint Alberta-Thy-29-94.}
setgray (DRAFT)

\begin{center}
{ \Large \bf
The Deuteron Spin Structure Functions

in the Bethe-Salpeter
Approach

and the Extraction of the Neutron

Structure Function $g_1^n(x)$\footnote{Talk given at the SPIN'94
International
Symposium, September 15-22, 1994, Bloomington, Indiana.}.

\phantom{.}}

{\large \noindent
  A.Yu. Umnikov, L.P. Kaptari$^\dagger$, K.Yu. Kazakov$^\ddagger$\\
and F.C. Khanna\\[2mm]
{\normalsize \em
 Theoretical Physics Institute, Physics Department,
 University of
 Alberta, \\Edmonton,
  Alberta T6G 2J1,\\ and
 TRIUMF, 4004 Wesbrook Mall, Vancouver, B.C. V6T 2A3,
 Canada}\\[1mm]
{\normalsize \em
 $^\dagger$Bogoliubov's Laboratory of Physics, JINR,
Dubna, 141980 Russia}\\[1mm]
{\normalsize\em
 $^\ddagger$Far Eastern State University, Vladivostok, 690000 Russia}
}
 \end{center}

\begin{abstract}

 The nuclear effects in the spin-dependent
 structure functions $g_1^D$ and $b_2^D$ are calculated
 in the relativistic approach  based on the Bethe-Salpeter
 equation with a realistic meson-exchange potential.
 The results of calculations are compared with the
 non-relativistic calculations. The problem of  extraction
 of the neutron spin structure function, $g_1^n$, from the
 deuteron data is discussed.
\end{abstract}

\vskip .3cm
\centerline{\large \bf 1. PRELIMINARIES}
\vskip .2cm

 In this talk we  present a
 relativistic approach to the deep
 inelastic lepton scattering on the deuteron based on the
 Bethe-Salpeter (BS)
 equation within a realistic meson-exchange model.
 The method is developed in refs.~\cite{umkh,umkh1}  and
 now we apply it to the deep inelastic scattering with
  polarized particles. We calculate the leading twist
 spin-dependent structure functions (SF) of the deuteron,
 $g_1^D$ and $b_2^D$~\cite{jaffe}, and we discuss the extraction
  of the neutron SF $g_1^n$ from deuteron data.
 This investigation is partially motivated
 by a number of  existing and forthcoming
 experiments on the deep inelastic scattering
 of leptons by deuterons (SLAC,  CERN, DESY, CEBAF).
 A covariant theory of this process will
 be useful  in the  analysis of the experimental data.

\vskip .3cm
\begin{center}
{\large \bf 2. THE BETHE-SALPETER AMPLITUDE FOR THE DEUTERON}
\end{center}

An accurate description
of both the $NN$-interaction at
 energies up to $\sim$1 GeV, and the basic
 properties of the deuteron, can be provided
within the
 meson-nucleon theory~\cite{tjond,gross}.
 The covariant description is
 based on the BS equation or its various
 approximations.  We use
 the ladder approximation for the kernel
of the
  BS equation~\cite{tjond}:
  {\small
\begin{eqnarray}
\Phi(p, P_D) =
i \hat S(p_1)\cdot\hat S(p_2) \cdot
\sum_{B} \int \frac{d^4p'}{(2\pi)^4}
\cdot
\frac{g^2_B \Gamma_{B}^{(1)}
\otimes
\Gamma_{B}^{(2)}}{(p-p')^2-\mu_B^2}
\cdot \Phi(p', P_D),
\label{sseq}
\end{eqnarray}
}
where  $\mu_B$ is the mass
 of meson $B$;
$\Gamma_{B}$ is the meson-nucleon vertex,
corresponding to the meson $B$, $\hat S(p)
= (\hat p - m)^{-1}$,
$m$ is the nucleon mass. This equation is solved, using
the technique described in ref.~\cite{umkh}.

The meson parameters, such as
masses, coupling constants, cut-off
parameters
are taken similar to those
in ref.~\cite{tjond},
with a minor adjustment
of the coupling constant of the scalar
$\sigma$-meson
 so as to provide a numerical solution
of the BS
equation.
 All parameters are presented
in Table 1, where coupling constants are
shown in accordance with
our definition of the meson-nucleon
form-factors,
$F_B(k) = (\Lambda^2 - \mu_B^2)/(\Lambda^2 - k^2)$.

The  deuteron amplitude  is normalized
by using the conserved vector current:
\vspace*{-5mm}
\begin{eqnarray}
J_{\mu}(0) = \langle P_D \mid
 \bar \psi \gamma _\mu \psi
 \mid P_D \rangle = 2  P_{D\mu} .
\label{curr}
\end{eqnarray}

For the analysis of the
processes with the polarized deuterons
an important characteristic of the deuteron
wave functions is the probability of the $D-$wave,

\hspace*{1cm}{\sf{\bf Table 1.} Parameters of the model}
\begin{center}
\begin{tabular}{|c|c|c|c|c|}
\hline
{\bf meson}&{\bf coupling constants}&{\bf mass}&{\bf cut-off}&{\bf
isospin} \\
 B   & $g^2_B/(4\pi); [g_v/g_t]$ &$\mu_B$, GeV  &$\Lambda$,GeV &
\\
 \hline
\hline
 $\sigma$ & 12.2 & 0.571 & 1.29 & 0  \\
\hline
 $\delta$ & 1.6 & 0.961 & 1.29  & 1 \\
\hline
 $\pi$ & 14.5 & 0.139 & 1.29   & 1\\
\hline
 $\eta$ & 4.5 & 0.549 & 1.29  & 0 \\
\hline
 $\omega$ & 27.0; [0] & 0.783 & 1.29  & 0 \\
\hline
 $\rho$ &  1.0; [6] & 0.764 & 1.29 & 1  \\
\hline
\hline
 \multicolumn{5}{|c|}{$m = 0.939$ GeV, $\epsilon_D = -2.225$ MeV}\\
\hline
\end{tabular}
\end{center}

\noindent
 ${\cal P}_D$,
 which varies
in the range $3-6 \%$ for realistic potentials.
Since the components of the BS amplitude do not have a direct
probability
interpretation, we consider the  matrix element of the axial
current on the deuteron state with
the total momentum projection $M=1$:
\begin{eqnarray}
J_{\mu}^5(0)= \langle P_D \mid
 \bar \psi \gamma_5\gamma _\mu \psi
 \mid P_D \rangle_{M=1} \label{curr50} \\
J_{3}^5(0)\equiv \int \limits_{0}^{\infty}
n_{spin}^{BS}(|{\bf p}|)|{\bf p}|^2 d|{\bf p}|,
\label{curr5}
\end{eqnarray}
which corresponds in the non-relativistic limit to the mean value of
the
spin projection:
\begin{eqnarray}
\langle P_D \mid
 \sigma_3
 \mid P_D \rangle_{M=1} &=& \int \limits_{0}^{\infty}
n_{spin}^{n.r.}(|{\bf p}|)
|{\bf p}|^2 d|{\bf p}|= 1 -\frac{3}{2}{\cal P}_D,  \nonumber\\
n_{spin}^{n.r.}(|{\bf p}|)&=& u^2(|{\bf p}|)
 -\frac{1}{2} w^2(|{\bf p}|).
\label{charge5}
\end{eqnarray}

Our calculation of the matrix element (\ref{curr5}) gives
$J_3^5(0) = 0.922$, from which we get
 an approximate estimate of
${\cal P}_D^{BS} \approx 5.2 \%$ in reasonable
agreement with the estimate ${\cal P}_D^{BS} = 4.8 \%$ from
ref.~\cite{tjond}
(compare, also, with ${\cal P}_D=4.3\%$ and ${\cal P}_D=5.9\%$
for Bonn and Paris potentials, respectively). Note, that
accuracy of such a computations of the $D-$wave admixture
from the (\ref{curr5}) is of the order
of the $\sim \langle p^2 \rangle /m^2$,  i.e. $\sim 1\%$.

The matrix elements (\ref{curr}) and (\ref{curr50}) contain
the explicit integration over the relative 4-momentum. To compare our
results with the non-relativistic case, we define the charge density
$ n_{ch.}^{BS}(|{\bf p}|)$ in the deuteron
as a primitive function in (\ref{curr}) and (\ref{curr5})
 prior
the integration on the modulus of the 3-momentum:
\begin{eqnarray}
 \langle P_D \mid
  \bar  \psi \gamma _\mu \psi
 \mid P_D \rangle \equiv \int \limits_{0}^{\infty}
n_{ch.}^{BS}(|{\bf p}|)|{\bf p}|^2 d|{\bf p}| ,
\label{charge}
\end{eqnarray}
which corresponds to the non-relativistic charge density:
\begin{eqnarray}
n_{ch.}^{n.r.}(|{\bf p}|) = u^2(|{\bf p}|)
 + w^2(|{\bf p}|),
\label{charge1}
\end{eqnarray}
where $u(|{\bf p}|)$ and $w(|{\bf p}|)$
 are $S-$ and $D-$waves components of the
deuteron wave function. Similarly the spin density
$ n_{spin}^{BS}(|{\bf p}|)$ is defined from eq. (\ref{curr5}).
The charge and spin densities are compared with
those obtained with wave functions of the Paris~\cite{paris}
 and Bonn~\cite{bonn} potentials
in Fig. 1 and 2, respectively.
All calculations are in a reasonable agreement in the
non-relativistic
region, up to $|{\bf p}| \sim m$.

\vskip .5cm
\begin{center}
{\large \bf 3. THE DEEP INELASTIC SCATTERING ON THE POLARIZED
DEUTERON}
\end{center}
\vskip .2cm

 To calculate the structure functions (SF)
 of the deuteron we use  the OPE
 method within the effective
 meson-nucleon theory~\cite{levin,physlet,umkh}. This method allows
us
to calculate SF in terms of the Wick rotated BS
amplitude~\cite{umkh,umkh1}. Neglecting possible "off-mass-shell"
corrections to the deuteron SF~\cite{thomof,weisof},  we
obtain the deuteron SF in the convolution
form.
For two leading twist spin-dependent
SF of the deuteron,
$g_1^D$ and $b_2^D$,~\cite{jaffe} we have:
\begin{eqnarray}
&& g_1^D(x) = \int \limits_{0}^{1} f^{N/D}_5(\xi) g_1^N
(x/\xi)\frac{d\xi}{\xi},
\label{conv1}\\
&& b_2^D(x) = \int \limits_{0}^{1} \Delta f^{N/D}(\xi) F_2^N
(x/\xi)d\xi,
\label{conv2}
\end{eqnarray}
where $g_1^N$ and $F_2^N$ are isoscalar nucleon SF.
The moments of the effective distribution functions,
$f^{N/D}_5$ and $\Delta f^{N/D}$ are the matrix elements
of the leading twist operators on the deuteron states:
\begin{eqnarray}
&&\!\!\!\!\!\!\!\!\!\!\!\!
\mu_n(f) = \int\limits_0^1 x^{n-1} f(x) dx, \label{mom} \\[2mm]
&&\!\!\!\!\!\!\!\!\!\!\!\!
\mu_n(f^{N/D}_5) = \left.\frac{1}{2M_D^n}\int \frac{d^4p}{(2\pi)^4}
  p_{1+}^{n-1}
\left [{\bar \Phi_{M}(p)}
 \gamma_+^{(1)}\gamma_5^{(1)}(p_2\gamma^{(2)}-m)
\Phi_{M}(p)\right ] \right |_{M=1},
\label{mom5}\\[4mm]
&&\!\!\!\!\!\!\!\!\!\!\!\!
\mu_n(\Delta f^{N/D}) = \frac{1}{2M_D^n}\int \frac{d^4p}{(2\pi)^4}
p_{1+}  ^{n-1}\label{momdel}\\
&&\quad\quad\quad\quad\quad\quad
\left \{ \left.\phantom{\int}\!\!\!\!\!\!\left [
{\bar \Phi_{M}(p)}
 \gamma_+^{(1)}(p_2\gamma^{(2)}-m)
\Phi_{M}(p)\right ] \right |_{M=0} -
\left.\phantom{\int}\!\!\!\!\!\!\left [\phantom{\bar \Phi_{M}}
\!\!\!\!\!\!\ldots\right ] \right |_{M=1} \right\},
\nonumber
\end{eqnarray}
where  the kinematical variables in the
rest frame are defined by
 \begin{eqnarray}
p = (p_0,{\bf p}),
\quad
P_D=(M_D, {\bf 0}), \quad
p_1 = \frac{P_D}{2} + p, \quad p_2
= \frac{P_D}{2} - p,
\label{kin}
\end{eqnarray}
where $M_D$ is the deuteron mass, $p_+ = p_0+p_3$,
$\gamma_+ = \gamma_0+\gamma_3$
 and we use deep inelastic kinematics:
$pq \approx q_0 (p_0+p_3)$.

 The explicit form of the effective distribution functions,
$f^{N/D}_5$ and $\Delta f^{N/D}$ is defined by the inverse Mellin
transform
of (\ref{mom5}) and (\ref{momdel}).
 These distributions
 satisfy  normalization conditions:
 \begin{eqnarray}
\int \limits_{0}^{1} f^{N/D}_5 (\xi) d\xi  = J_3^5(0),
\quad \quad\int \limits_{0}^{1} \Delta f^{N/D}(\xi) d\xi  = 0.
\label{spinnor}
\end{eqnarray}

Using the  distribution functions, $f^{N/D}_5$ and $\Delta f^{N/D}$,
and realistic parametrizations of the nucleon SF, $g_1^N$~\cite{shaf}
and
$F_2^N$~\cite{physlet}, we calculate the deuteron SF $g_1^D$ and
$b_2^D$.
The results of calculation are presented on Figs. 3 and 4.
In Fig. 3
nuclear effects in $g_1^D$ calculated within BS approach (solid
curve)
 are compared
with non-relativistic calculation~\cite{g1germania,g1italia}
with Bonn wave function. The non-relativistic calculation including
the
Fermi motion of polarized nucleons and binding effects (dashed curve)
is in reasonable agreement with the relativistic calculations (solid
curve).
At the same time both these curves differ from non-relativistic
impulse approximation (dotted), which does not include binding
effects.
The SF $b_2^D$ calculated in relativistic and non-relativistic
approaches is presented in Fig. 4. Similar to the case of the
$g_1^D$,
there is no special relativistic effect in $b_2^D$. The models are
slightly
distinguishable in view of different admixture of $D$-wave and
minor variations of the nucleon momentum distributions.
\newpage
\vskip .5cm
\begin{center}
{\large \bf 4. ON THE EXTRACTION OF THE NEUTRON SF.}
\end{center}
\vskip .2cm

Mathematically the problem of extraction of the neutron
SF from the deuteron data is formulated as a problem to solve
the inhomogeneous integral equation (\ref{conv1})
for the neutron SF with a model kernel $f^{N/D}_5$
and experimentally measured left hand side\footnote{Depending
of the model, some
additive corrections could be taken into account~\cite{unfo}.},
$g_1^D$.

Recently we proposed a method
to extract the
neutron SF from the deuteron data
within any
model, giving deuteron SF in the form
of a "convolution integral plus/minus additive
corrections"~\cite{unfo}.
The principal advantages of the method, compared with
the smearing factor method, are the following.
(i) Only analyticity of the SF need
be assumed, (ii) the
method allows us to elaborate on the spin-dependent SF,
where the traditional smearing factor method does not work.

As an example, we apply our method to the SMC data~\cite{smc}.
The resulting
nucleon functions are presented in fig. 5 (solid curves).
It is clear that due to a singular behavior of the ratio
$g_1^D/g^N_1$ the
nucleon SF  $g^N_1$ cannot be obtained by
the smearing factor method. As a more practical result we
can report that even in the experimentally restricted region  of
$x$ the following integral relation is valid
with high accuracy:
 \begin{equation}
 \int \limits_{0.006}^{0.6}  g_2^D (x) dx \cong
(1-\frac{3}{2}{\cal P}_D) \cdot \int \limits_{0.006}^{0.6}  g_2^N(x)
dx,
 \label{bsr}
 \end{equation}

\vspace*{2mm}
\centerline{\large \bf 5. CONCLUSIONS}
\vspace*{2mm}

  We have presented a description
 of  the spin-dependent deep inelastic
 lepton-deuteron scattering based on the Bethe-Salpeter formalism
 within an effective meson-nucleon theory. In particular,
 \begin{enumerate}
 \item The spinor-spinor Bethe-Salpeter equation for the deuteron
 is solved in the ladder approximation
 for a realistic meson exchange potential.
 \item The leading twist spin-dependent structure functions,
 $g_1^D$ and $b_2^D$, of the deuteron are calculated
  in terms of the
 Bethe-Salpeter
 amplitude.
 \item Numerical results of the calculations of the
 deuteron structure functions $g_1^D$ and $b_2^D$ in the
Bethe-Salpeter
 formalism are presented. It is found
 that results are in qualitative agreement
with previous non-relativistic
 calculation.
 \item  A method to extract  the neutron structure function,
 $g_1^n$, from the deuteron and proton data is suggested.
 \end{enumerate}

 The reasonable quantitative agreement of the presented
 calculations of
 the deuteron SF at $x < 1$
 in the non-relativistic
 and relativistic approaches confirms the expectation that
 these approaches have to give similar results
 within the boundaries of validity of the
 non-relativistic
 approximation. However, it does not imply that the relativistic
 effects in the deuteron SF are
 negligible in general.
 It only shows that in a slightly relativistic system
such as the
 deuteron  we should
 find {\em special} kinematic conditions of the experiment to
 display the relativistic effects.
 We could expect
 non-trivial relativistic phenomena at $x > 1$,
  where the
precise evaluation of the SF
is important for QCD analysis of the
experimental data. The behavior of the  nuclear SF
$\sim (1-x_N)^\gamma$ as $x_N \to 1$
may lead to  errors.

\vskip 1cm

 {\large \bf Figure captions:}
{\footnotesize \sf

{\bf Figure 1.}
The charge density calculated in different models (see text).

{\bf Figure 2.}
 The spin density calculated in different models (see text).

{\bf Figure 3.} The ratio of the deuteron and nucleon SF, {\rm
$g_1^D/g^N_1$}.

{\bf
Figure 4.} The deuteron SF {\rm $b_2$} calculated in different models
(see text)

{\bf Figure 5.} The SF {\rm $g_1$} of deuteron (dashed lines) and
nucleon (solid lines) for two different parametrizations of data.

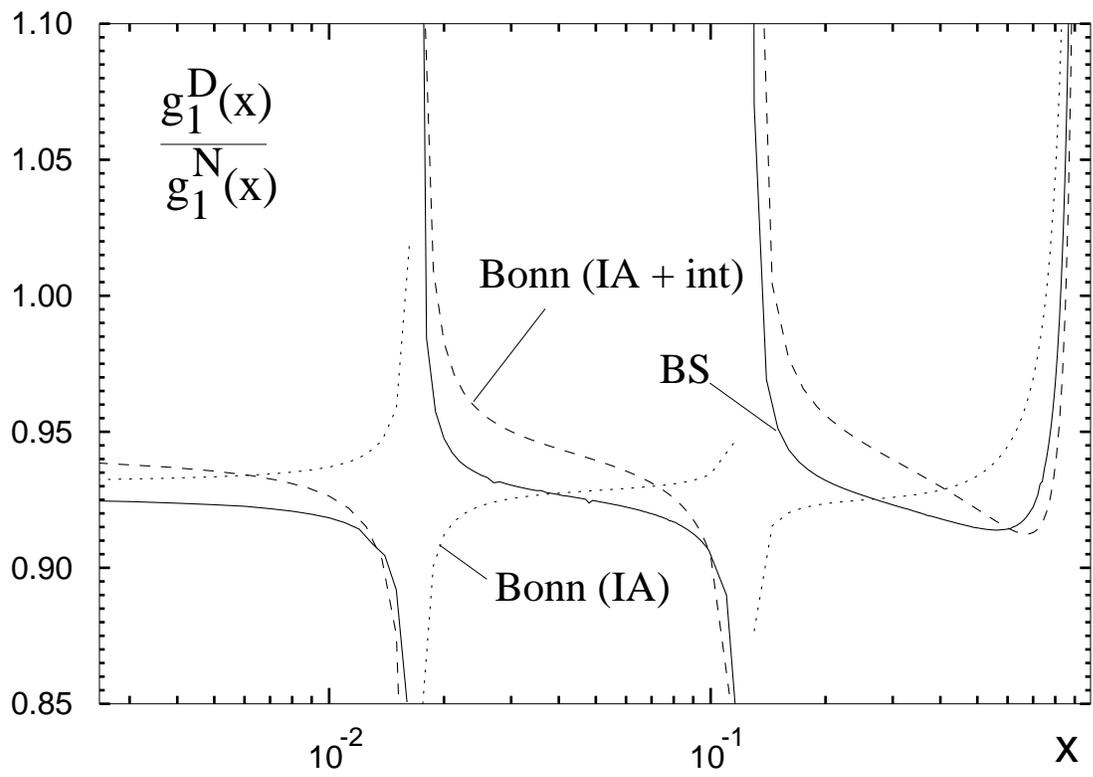
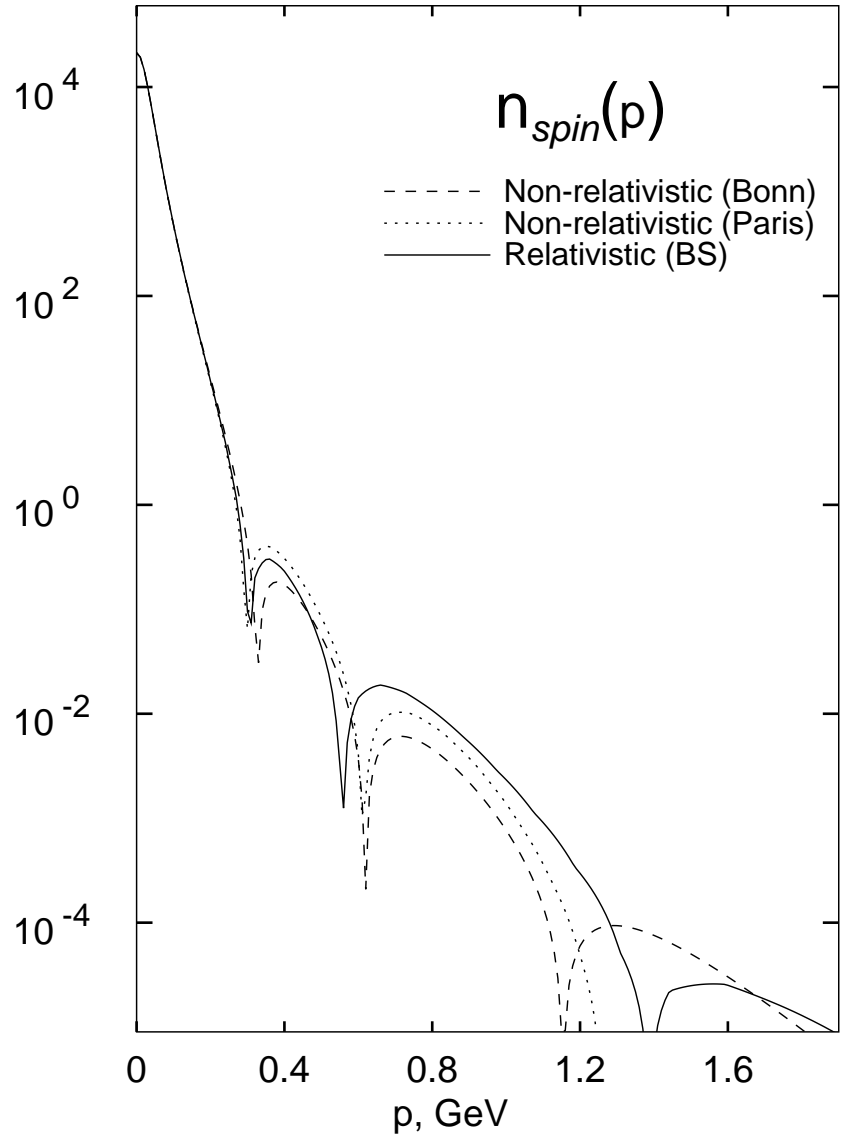
\end{document}